\documentclass[final,3p,times]{elsarticle}

\usepackage{mathtools}

\usepackage{amssymb}
\usepackage{subcaption}

\usepackage{amsthm}
\usepackage{amsmath}
\usepackage{enumitem}
\usepackage{booktabs}
\usepackage{multirow}
\usepackage{lscape}
\usepackage{tabu}
\usepackage{threeparttable}
\usepackage{natbib}

\usepackage{tikz}
\usetikzlibrary{shapes.geometric, arrows}
\tikzstyle{startstop} = [rectangle, rounded corners, minimum width=3cm, minimum height=1cm,text centered, draw=black, fill=red!30]
\tikzstyle{input} = [trapezium, trapezium left angle=60,trapezium right angle=120,minimum width=1cm, minimum height=1cm, text centered, draw=black, fill=blue!5, text width=2cm]
\tikzstyle{output} = [trapezium, trapezium left angle=120,trapezium right angle=60, minimum width=1cm, minimum height=1cm, text centered, draw=black, fill=blue!20, text width=3cm]
\tikzstyle{NN} = [trapezium, minimum width=4.5cm, minimum height=1cm, text centered, draw=black, fill=gray!40]

\tikzstyle{process} = [rectangle, rounded corners, minimum width=4.5cm, minimum height=1cm, text centered, draw=black, fill=gray!20]
\tikzstyle{decision} = [diamond, minimum width=3cm, minimum height=1cm, text centered, draw=black, fill=green!5]
\tikzstyle{data} = [rectangle, rounded corners, minimum width=1cm, minimum height=1cm,text centered, draw=black, fill=red!30]
\tikzstyle{arrow} = [thick,->,>=stealth]
\tikzstyle{arrow2} = [dashed,->,>=stealth]

\usepackage[pagewise]{lineno}

\usepackage{setspace}

\journal{Energy Efficiency Journal}

\begin{document}

\begin{doublespacing}
\begin{frontmatter}
    
    
    
    \title{\textbf{Investigating Opinion Dynamics Models in Agent-Based Simulation of Energy Eco-Feedback Programs}}

    \author{Mohammad Zarei $^1$}
    \author{Mojtaba Maghrebi $^2$}

    \address{$^1$ Ph.D. Candidate, Department of Civil and Environmental Engineering, University of Waterloo, 200 University Ave., Waterloo, ON N2L3G1, Canada (corresponding author). E-mail: mzarei@uwaterloo.ca}
    \address{$^2$ Associate Professor, Faculty of Engineering, Dept. of Civil Engineering, Ferdowsi Univ. of Mashhad, Azadi Square, 9177948974 Mashhad, Iran. E-mail: mojtabamaghrebi@ferdowsi.um.ac.ir }

    \begin{abstract}
        According to research, reducing consumer energy demand through behavioural interventions is an important factor of efforts to reduce greenhouse gas emissions and climate change.On this basis, feedback interventions that make energy consumption and conservation efforts apparent are seen as a feasible method for increasing energy-saving habits. Simulation techniques provide a convenient and cost-effective tool for examining the parameters that may affect the amount of energy saved as a result of such interventions. However, constructing a reliable model that accurately represents real-world processes is a significant issue.  Five Opinion Dynamic (OD) models that depict how opinion change occurs among individuals' interactions are investigated in this paper, and a Revised OD (ROD) model is suggested to develop more efficient eco-feedback simulation models. The results show that the influence condition and the weight-factor of connected opinions have a substantial impact on the accuracy of simulation outputs when compared to field experiment reports. As a result, ROD has been proposed for eco-feedback program simulations, as it provides the nearest approximation to the field data.

    \end{abstract}
    
    \begin{keyword}
       Opinion Dynamics \sep Energy Efficiency Interventions \sep Social Network \sep Agent-Based Simulation \sep Eco-Feedback Program  
    \end{keyword}

\end{frontmatter}
\end{doublespacing}


\section{Introduction}
\label{section:intro}

The growing industrial and residential carbon footprints contribute to the ongoing climate change process.Regulatory legislation can limit the contribution of industry to greenhouse gas emissions. However, the energy consumption of buildings, which accounts for up to 40\% of global energy consumption in developed countries and roughly the same percentage of gas-emission production is difficult to regulate without causing dissatisfaction among residents \cite{perez2008review, chen2021impacts}. As a result, building occupants can be identified as an important target group for energy-saving goals. In other words, by focusing on energy-related behaviours in buildings, significant energy savings could be realised \cite{deumling2019everyone,karjalainen2016should}. It is worth noting here a phenomenon known as the "take back" or "rebound" effect, in which households tend to increase their energy consumption following physical promotions, making improving household energy behaviour a higher priority than technical improvements \cite{adha2021rebound}. As a result, physical improvement without behavioural interventions will be less effective than expected \cite{masoso2010dark,gynther2012evaluation}.

In general, occupant energy behaviour is a dynamic attribute that can be influenced by social interactions \cite{gynther2012evaluation, xie2021exploring, mukai2022effect} as well as energy-saving interventions \cite{abrahamse2007effect, jain2013can}. Among different types of occupancy interventions for energy conservation (such as goal setting \cite{abrahamse2007effect}, commitment \cite{hu2020enhancing}, and workshops \cite{mane2014analysis}), the feedback method has drawn the most attention from scholars \cite{abrahamse2007effect, fischer2008feedback, fischer2007influencing}. In a typical feedback program, occupants are given information about their energy consumption and can see how effective their efforts to conserve energy are. As a result, feedback interventions could address the invisibility of energy consumption and encourage consumers to engage in more sustainable behaviours \cite{guerreiro2015making}.

Feedback results could be improved in a variety of ways, including increasing the frequency of feedback, providing a time-, room-, or application-specific breakdown, improving the visual design, or adding extra details, such as time series, average comparisons, or information about environmental impact \cite{fischer2008feedback}. However, using eco-feedback programs to leverage social influences on occupant energy behaviour has been shown to be more efficient \cite{deumling2019everyone,jain2013can}. In this approach, occupants can compare their energy usage and their energy saving amount with their peers in their social network. 

In addition to empirical studies on energy feedback methods, several efforts have been made to simulate the impact of energy feedback on occupant energy behaviour \cite{zarei2020improving, zarei2020targeted, azar2017multilayer, anderson2014impact}. It is critical to create a reliable simulation model that can provide a cost-effective opportunity to study the effective parameters in feedback interventions. It is critical to create a reliable simulation model that can provide a cost-effective opportunity to study the effective parameters in feedback interventions. Using such models instead of empirical investigations could not only save time and money, but also allow researchers to investigate the impact of a wide range of factors on the efficiency of a feedback programs and provide practical recommendations. For example, in a simulation environment, various combinations of target participants for a feedback program can be modeled, and the characteristics of the best combination, which can lead to more energy savings, can be studied \cite{zarei2020improving, zarei2020targeted}. 

Feedback simulation studies have been commonly employed Agent-Based Modeling (ABM) as a powerful approach in simulating complex systems of autonomous agents \cite{macal2005tutorial} integrating with Opinion Dynamics (OD) models, which are mathematical models for describing the process of attitude/opinion change among individuals \cite{xia2011opinion}.This study will use ABM to investigate various types of OD models and compare them based on the accuracy of energy saving prediction for eco-feedback simulation in the residential context. 

The layout of the paper is as follows. The following sections give a brief background on energy feedback interventions, Agent-Based Modeling (ABM) and Opinion Dynamics (OD) models. Then in the methodology section the three-step process, by which different OD models have been evaluated are explained. Finally, the simulations results and sensitivity analysis are presented, discussed and concluded.

\section{Agent-Based Modeling (ABM)}
ABM is a method of representing autonomous entities ('agents') and simulating the outcomes of these intelligent agents' interactions within an environment through the implementation of rule-based decisions that result in a variety of potential outcomes; agents include people, businesses, animals, organisations, and equipment \cite{grimm2006standard, macal2016everything}. ABM has already been investigated to aid energy simulation in building energy studies \cite{zarei2020improving, zarei2020targeted, azar2017multilayer}. Azar et al \cite{azar2012agent} proposed an approach for energy estimation by considering different and dynamic energy behavior among building occupants. This study has been continued in evaluating saving potential from occupancy interventions in typical commercial buildings \cite{azar2014framework} and energy feedback methods for groups of buildings \cite{azar2017multilayer}. The findings of the mentioned studies indicate that combining occupants’ energy behavior and the social network influence with energy simulation models could increase the effectiveness of occupancy interventions models. 

Another study used Bass diffusion theory in an agent-based environment to represent the spread of energy-saving policies among occupants, as well as the associated impact on energy consumption and emission production \cite{bastani2016application}. According to simulation results, it has been concluded that the occupants' word-of-mouth effect had a strong influence on persuading them to save energy. This is yet another example of the significance of eco-feedback interventions.

In a more recent study using ABM for simulating energy efficiency programs, Zarei and Maghrebi \cite{zarei2020improving} showed showed that considering characteristics of participants can improve energy saving outcomes. Consequently, they proposed an a Genetic Agent-Based (GAB) method to enhance the return of energy saving programs by simulating social network and energy behavior attributes and finding the best participants among a target community \cite{zarei2020targeted}. 

In all studies related to the simulation of energy consumption, specifically through occupancy interventions such as eco-feedback, the main challenge is to mathematically describe the dynamics of energy related behaviors among occupants. Using Opinion Dynamics (OD) Models, which are discussed in the following section, is a common way to overcome this challenge.

\section{Opinion Dynamics (OD) Models}
\label{section:OD}
In the context of sustainability issues, the aggregate sum of individual opinions that lead to specific actions has a direct impact on both the local (air or water quality, noise disturbance, etc.) and global environments (climate change, resource scarcity etc.). Fortunately, opinions could be revised \cite{duggins2014psychologically}, which is a complex process affected by interplay of different elements including the individual predisposition, the influence of positive and negative peer interaction (social networks playing a crucial role in this respect), the information each individual is exposed to, and many others \cite{xia2011opinion}. In this regard, models of Opinion Dynamics (OD) can be applied as a quantitative tool that simulate the psychological and social aspects of opinion change to investigate mechanisms driving citizens’ environmental awareness. This awareness could eventually lead to more sustainable behaviours.

There are several types of OD models in the literature. In the following, five common OD models that consider continuous values for opinions are presented. Each model, mathematically describe the process of changing opinion (e.g. $x_i$) through interactions with other connected opinions (e.g. $x_j$):

\begin{itemize}
    \item \textit{Deffuant-Weisbuch (DW)} \cite{deffuant2001mixing}:
    \begin{equation}
        |x_i^t-x_j^t|<d \rightarrow x^{t+1}_i = x_i^t + \mu \times (x^t_j - x_i^t)
    \end{equation}
    
    \item \textit{Jager-Amblard (JA) } \cite{jager2005uniformity}:
    \begin{equation}
    x^{t+1}_i=
        \begin{cases}
        x_i^t + \mu \times (x^t_j - x_i^t),\,\, \text{ if } |x_i^t-x_j^t|<u_i\\
        x^{t+1}_i = x_i^t + \mu \times (x^t_i - x_j^t),\,\, \text{ if } |x_i^t-x_j^t|<t_i 
        \end{cases}
    \end{equation}
    
    \item \textit{Social Influence Network (SIN) } \cite{friedkin2001norm}:
    \begin{equation}
        x^{t+1}_i = (1-a_i)\times x_i^t + a_i \times \sum^n_{j=1} w_{ij}^{SIN}\times x_j^t
    \end{equation}
    
    \item \textit{Relative Agreement (RA)} \cite{amblard2004role}:
    \begin{equation}
            x^{t+1}_i= x_i^t + \mu \times (x^t_j - x_i^t) \times \left(\frac{h_{ij}}{u_i^t}-1\right),\,\, \text{ if } h_{ij}>u_i^t
    \end{equation}
    \begin{equation}
            u^{t+1}_i= u_i^t + \mu \times (u^t_j - u_i^t) \times \left(\frac{h_{ij}}{u_i^t}-1\right),\,\, \text{ if } h_{ij}>u_i^t 
    \end{equation}
    \begin{equation}
        h_{ij}= \min(x_i^t+u_i^t, x_j^t+u_j^t) - \max(x_i^t-u_i^t, x_j^t-u_j^t))
    \end{equation}
    
    \item \textit{Influence, Susceptibility, and Conformity (ISC) } \cite{duggins2014psychologically}:
    \begin{equation}
        x^{t+1}_i = x_i^t + \frac{1}{k_i} \times \frac{\sum^n_{j=1} w_{ij}^{ISC} (\hat{x}_j^t-x_i^t)}{\sum^n_{j=1} w_{ij}^{ISC}}
    \end{equation}
    \begin{equation}
        \hat{x}^{t}_j = x_j^t + \frac{c_j}{k_j} \times \frac{\sum^n_{k=1} (\hat{x}_k^t-x_j^t)}{n}
    \end{equation}
    \begin{equation}
        w_{ij}^{ISC} = 1-2\times \frac{|x_i^t-x_j^t|}{x_{max}}
    \end{equation}
    
\end{itemize}

\noindent where 

\begin{itemize}
    \item[] $x_i^t$: Numerical value assigned for the opinion of person $i$ at $t$
    \item[] $x_i^(t+1)$: Numerical value assigned for opinion of person i at $t+1$
    \item[] $x_j^t$: Numerical value assigned for opinion of person  $j$ at $t$
    \item[] $d$: Opinion difference threshold
    \item[] $\mu$: The strength of influence in DW and rate of the dynamics in RA
    \item[] $u_i$: The latitude of acceptance
    \item[] $t_i$: The latitude of rejection
    \item[] $a_i$: Susceptibility of person $i$ to the influence of others
    \item[] $w_{ij}^{SIN}$: Influence strength of the $j$ on person $i$; the number of common social connections between agent $i$ and $j$
    \item[] $u_i^t$: Opinion uncertainty of person $i$ at $t$
    \item[] $c_j$: Inherent willingness of j to misrepresent his/her beliefs in social contexts in order to appear either normal or distinct
    \item[] $k_i$: Inherent commitment of $i$ to his/her current opinion
\end{itemize}

Except for SIN and ISC models, there is an influence condition in other OD models that numerically determine which opinions in the social network of agent $i$ could change his/her opinion (i.e. $x_i$). These models use the similarity of opinions as a required permit for opinion change. In DW model, an opinion difference threshold is defined, and in RA model the overlapping of opinions should be more than opinion uncertainty of agent $i$. Lastly, in JA model if the dissimilarity of opinions is greater than a certain amount (i.e. the latitude of rejection), their interaction will lead to the even more different opinions.

According to these OD models, after collecting all acceptable interactions based on influence conditions in each cycle, the opinion change occurs based on the given equation. The equation has two common components that connect the new opinion, $x_i^{(t+1)}$, to the previous opinion, $x_i^t$, and connected opinions, $x_j^t$:
\begin{itemize}
    \item \textit{Flexibility-factor} that determines how far an opinion could change through social interactions. This factor has been named as the strength of influence in DW and JA, susceptibility in SIN, rate of the dynamics in RA, and inherent commitment in ISC.
    \item \textit{Weight-factor} that represent which opinions have more influence on a particular agent. This parameter is considered equally for all opinions in DW model. In JA model, this factor could be equal to -1 or +1 based on comparison of opinion difference with the latitude of rejection and the latitude of acceptance. Other models have more complex definition for weight-factors as presented in the formulation.
\end{itemize}
	
Some of the models mentioned above have been used as the central logic of simulations in the area of occupant energy behaviour modelling. For example, Azar and Menassa used a RA model in their ABM to simulate discrete occupancy focused interventions as well as continuous peer interactions for the social subnetworks found in a typical commercial building in the United States \cite{azar2012agent}. Anderson et al use the SIN model to investigate the impact of social network type and structure on modelling normative energy use behaviour interventions in a similar study \cite{anderson2014impact}.
This study evaluates the accuracy of these OD models in simulating energy eco-feedback interventions in residential buildings. After comparing the model predictions for energy savings to the field experiment reports, a Revised OD (ROD) model for eco-feedback simulations is proposed. The following section describes the methodology used to analyse and compare the OD models.

\section{Methodology}

To address the research objective, which is evaluating different OD models in eco-feedback simulation, an agent-based environment is developed using Python 3.6. The simulation process consists of three main steps, generating agents (households), providing eco-feedback information for each agent, and estimating energy saving range.

\subsection{Generating Agents}

The model entities (agents) are a target community of connected households who have different Energy Index (EI) and specific position in the community social network. Energy Index (EI) represents household’s energy consumption behavior on a scale of 1 to 100, which 1 is the least energy consumer household and 100 is the highest consumer. This attribute for each agent is derived from a lognormal distribution ($\sigma=0.388,\mu=-0.924$), which is the best obtained fit among known restricted distributions to the energy use data acquired from a typical 100-unit residential building located in Mashhad, Iran \cite{zarei2020improving}. The result of Kolmogorov-Smirnov (KS) statistic is 0.032, which is less than the 0.05 common threshold confirming the goodness of fit. In this process each household’s EI is calculated as follows:

\begin{equation}
    EI_i = \frac{EU_i-EU_{min}}{EU_{max}-EU_{min}}
\end{equation}
\noindent where $EU_{max}$, $EU_{min}$, and $EU_i$ are the maximum annual electricity consumption per unit area, minimum annual electricity consumption per unit area and the annual electricity consumption per unit area for household $i$ respectively. 

\begin{figure}
    \centering
    \includegraphics[width=1\linewidth]{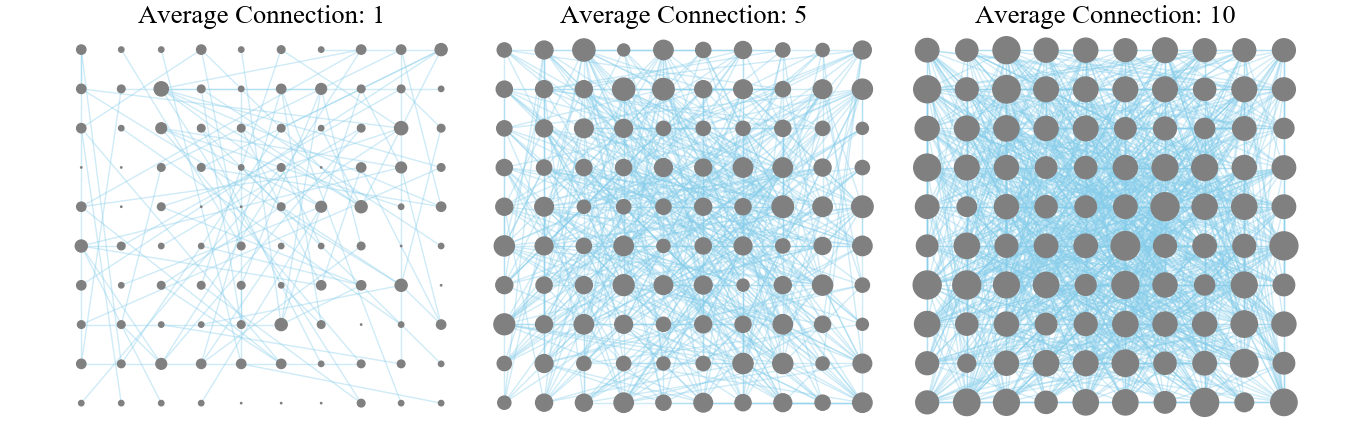}
    \caption{Random ER graphs with 100 nodes, and different ANC using Networkx library \cite{hagberg2008exploring}}
    \label{fig:network}
\end{figure}

As mentioned before, the simulation results of OD models have been compared to empirical experiments. Since there is lack of information about the structure of social networks in these experiments, Random ER (Erdos-Renyi) \cite{gomez2006scale} graph with a random average number of connections (ANC) is assumed for the social network of households (Figure \ref{fig:network}). Thus, the possible impact of social network structure has been minimized. However, the sensitivity of outputs to ANC has been analyzed in the last section. 

Random ER (Erdos-Renyi) graphs are generated by starting with a set of isolated nodes that are then paired with a uniform probability. Most nodes have the same number of connections in these networks and the degree distribution will be a Gaussian bell-shaped curve \cite{gomez2006scale}. To build ER graphs, Networkx \cite{hagberg2008exploring} as a Python-based package for anslysing complex network structures has been utilized. Figure \ref{fig:network} depicts three random ER graphs with 100 nodes that have one, five and ten as ANC respectively. The size of each node is related to the number of connections, which commonly referred as node degree.

\subsection{Energy Eco-Feedback Program}
In this step, energy eco-feedback is provided for agents so that their energy behavior could be modified accordingly. To represent this process mathematically, the mentioned OD models in section \ref{section:OD} are used. EI is assumed to play as the quantified opinion (i.e. $x_i$ in equations in Section \ref{section:OD}) of each agent towards energy conservation issues. The greater this attribute, the less concern for energy conservation. Furthermore, the agents are only given feedback on their lower energy consumption. During each simulation run, Energy Indexes (EI) of agents (i.e. households) will be revised according to Equation \ref{eq:EI change}, which is a general equation based on OD models:

\begin{equation}
    \label{eq:EI change}
    \hat{EI}_i = EI_i + \epsilon \times \frac{\sum^n_{j=1} w_{ij}\times (EI_j - EI_i)}{\sum^n_{j=1} w_{ij}}
\end{equation}

Where $\hat{EI}_i$ and $EI_i$  are denoted to Energy Index of agent $i$ after and before eco-feedback program. $\epsilon_i$ indicates the flexibility-factor of agent $i$. This attribute is generated from a normal distribution with a mean value of 0.5 and a standard deviation of 0.3. These values shows that people have different rates of behavior change and adoption to social norms. So that, distribution type and inputs can only make a relative change in system behavior \cite{zarei2020improving,zarei2020targeted,azar2017multilayer, anderson2014impact}. Finally, $w_{ij}$ reflects the weight-factor that represent the strength of relationship between agent $i$ and his/her connections. The influence conditions and weight-factor for OD models are considered as follows:

\begin{itemize}
    \item DW model: $d_i = EI_i \times \epsilon_i$, and $w_{ij} = 1$ 
    \item JA model: $u_i=t_i= EI_i\times \epsilon_i$, and $w_{ij} \in \{1,-1\}$
    \item SIN model: there is no influence condition, and $w_{ij}=c_{ij}$ is the number of common social connections between agent $i$ and $j$
    \item RA model: $u_i = \min(EI_i, 100-EI_i)$, and $w_{ij} = \frac{h_{ij}}{u_i}-1$
    \item ISC model: there is no influence condition, and $w_{ij} = 1- \frac{|EI_i-EI_j|}{50}$
\end{itemize}

In addition, we have proposed an revised OD model (ROD) with adjusted influence condition and weight factor as follows:

\begin{itemize}
    \item ROD model: $h_{ij} > EI_i \times \epsilon_i$, and $w_{ij} = \left(1- \frac{|EI_i-EI_j|}{50}\right)\times c_{ij}$
\end{itemize}

\noindent where $h_{ij}$ is computed based on RA model. The proposed ROD model can be seen as a hybrid form of all other OD models. The influence condition is derived from DW and RA, which limits the opinion changes from very different EIs. The weight-factor that is dependent on the proximity of EIs and social networks, is derived from ISC and SIN models.

\subsection{Energy Saving Estimation}
The change in the EI of agents during eco-feedback means that their energy consumption rate have been modified. As a result, the percentage of energy consumption reduction ($ER$) after providing feedbacks for all agents can be calculated as follows:
\begin{equation}
    \label{eq:ER}
    ER = \frac{\sum_{i=1}^n (\hat{EI_i}-EI_i)}{\sum_{i=1}^n EI_i} \times 100
\end{equation}

\noindent where $n$ denotes the number of agents, $EI_i$ and $\hat{EI}_i$ denote the Energy Index for agent $i$ before and after intervention, respectively.

After creating a random social network of agents and assigning their attributes (e.g., EI and $epsilon$), the energy consumption feedback of connected agents is provided to them (each agent could only see the lower consumers in his/her social network) in order to modify their energy behaviour (EI) based on the eco-feedback information, Equation \ref{eq:EI change}, and the parameters related to the corresponding OD model. The results of each OD model for modelling eco-feedback impacts on household energy behaviour are compared to field data in the following section.

\section{Simulation Results and Discussion}

According to research objective that is evaluation of OD models for eco-feedback simulation, an agent-based environment, which has been explained in the previous section, is programmed using Python 3.6. To have a benchmark, a wide range of observed energy savings feedback experiments extracted from feedback review studies. The results varied considerably, but mostly between 5\% to 12\% \cite{abrahamse2007effect, darby2006effectiveness,francisco2018occupant,fischer2008feedback,khosrowpour2018review, dougherty2016behavioral}. The following are the selected parameters for model initiation that have been set based on the case study settings and previous models feedback models \cite{zarei2020improving,zarei2020targeted, azar2012agent, azar2017multilayer, anderson2014impact}:

\begin{itemize}
    \item Number of agents is set to 100 to represent a typical residential complex
    \item EI and $\epsilon$ values are randomly drawn from $Lognormal(\mu=-0.924,\sigma = 0.388)$ and $Normal(\mu=0.5, \sigma = 0.3)$, respectively.
    \item Social network type is Random ER with Average number of connections (ANC) randomly from 1 to 10.
    \item The social network type is Random ER, and the average number of connections (ANC) is chosen at random from 1 to 10.
    \item The simulation is repeated 1000 times.
\end{itemize}

\begin{figure}
    \centering
    \includegraphics[width=0.5\linewidth]{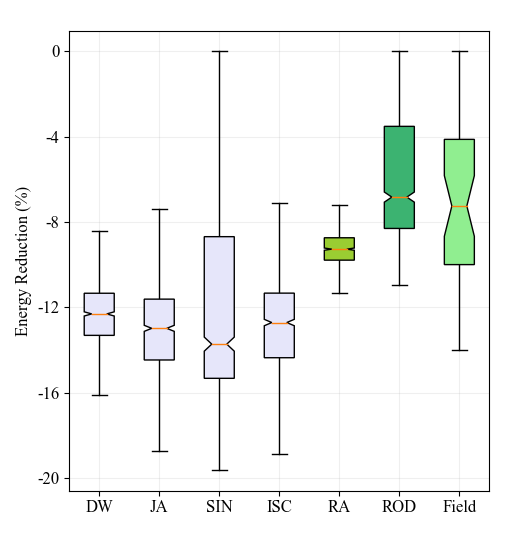}
    \caption{Energy reduction based on OD models versus empirical reports}
    \label{fig:results}
\end{figure}

Figure \ref{fig:results} illustrates the outputs of 1000 simulation runs for each OD model as a box-and-whisker diagrams comparing with field data. The vertical axis indicates the percentage of energy saving (i.e. ES), which is computed based on changes in the EI of all agents before and after the eco-feedback (Equation \ref{eq:ER}). The horizontal axis categorizes the results according to the OD model. The field experiment reports of energy saving are also given as the benchmark.

The first finding of Figure \ref{fig:results} is that different OD models can result in a wide range of energy savings via eco-feedback interventions. In other words, using the suitable OD model is critical for producing an accurate model. SIN and RA, for example, gave the broadest and narrowest ranges of energy reductions, respectively. This observation could be attributed to SIN's strong dependence on average number of connections (ANC) and the absence of influence condition, as opposed to RA's constrained influence condition.

It should be noted here that because the agents only get lower energy consumption of their peers as eco-feedback information, all of the OD models predict a negative total energy reduction, which is consistent with field evidence. The field experiment findings show a wide range of energy savings, ranging from 5 to 12 percent. Other OD types show higher amounts of energy reduction than field data, with the exception of the RA model, which fairly gives a close prediction about energy savings. Again, this could be due to the RA model's stricter influence condition, which restricts the impact of very low consumption feedback on extremely high consumers.

Given that all OD models follow Equation \ref{eq:EI change} change, the varying results of this simulation demonstrate the significance of influence condition and weight-factors. As a result, in order to generate valid models, simulation studies should pay more attention to behaviour model selection. In this aspect, ROD produces the most similar results to field data in terms of average and range of energy reductions. The closeness of energy behaviour (derived from RA) and the closeness of social ties (derived by SIN) are both considered in the computation of weight-factors in ROD. Furthermore, the influence condition is determined by the susceptibility of agents (derived from RA and DW models).

\begin{figure}
    \centering
    \includegraphics[width=0.8\linewidth]{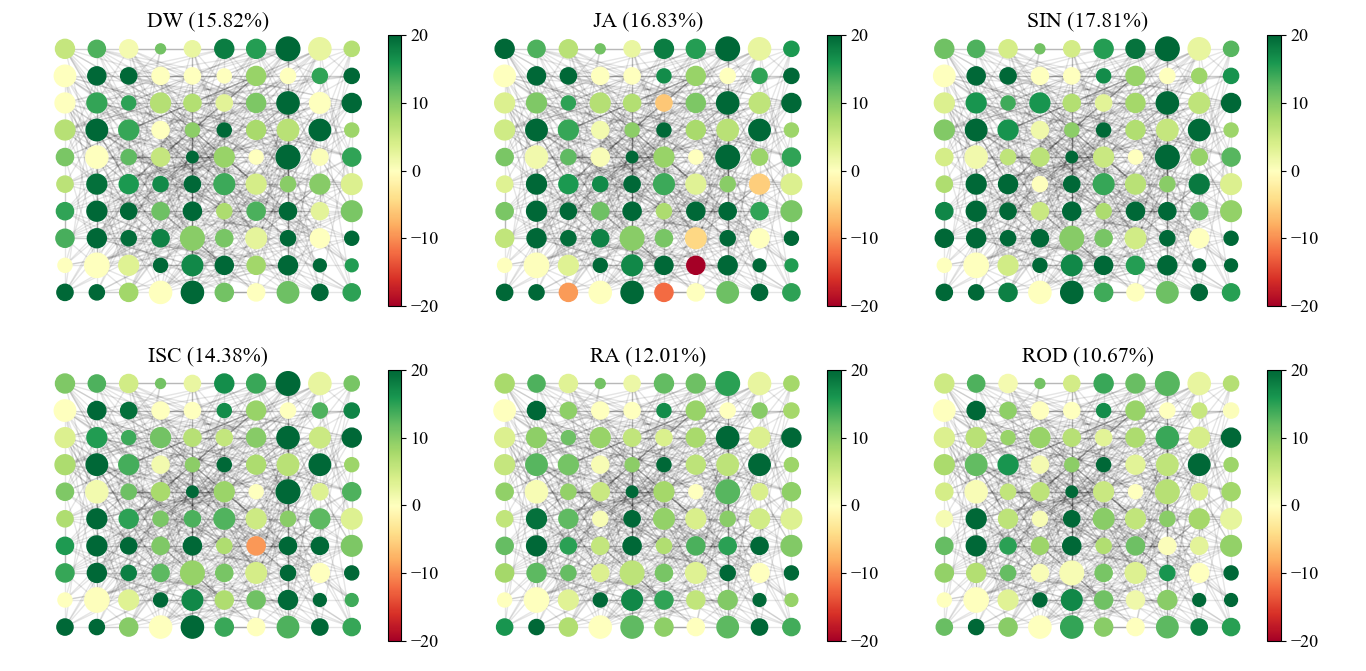}
    \caption{A snapshot from a one simulation run. Colors illustrate ER for each agent.}
    \label{fig:snap}
\end{figure}

Figure \ref{fig:snap} shows a snapshot from a simulation run to provide further insight at the micro-level. This figure depicts the energy saving projections of OD models for each agent in a 100-agent community with an ANC of five (see Fig. 3). For each OD model, the average energy savings from all agents are also determined.

Each node (agent) has a different size and colour. The colour represents the amount of energy saved on a scale of -20\% (red spectrum) to 20\% (green spectrum), and the size is proportional to the node degree (connections number of each agent).

With a quick glance at Figure \ref{fig:snap}, it is clear that the majority of the agents have positive energy savings (reducing their energy consumption as expected). However, there are few red nodes, indicating that eco-feedback is having a negative impact on these agents (energy consumption increased). The explanation for this observation, which arises solely in the JA and ISC models, is the weight-factor calculation approach. According to Table 2, weight-factors for the JA and ISC models could be negative if the two opinions are diametrically opposed. As a result, it is possible that eco-feedback information will have a negative impact.

\subsection{Sensitivity Analysis}

\begin{figure}
    \centering
    \includegraphics[width=0.8\linewidth]{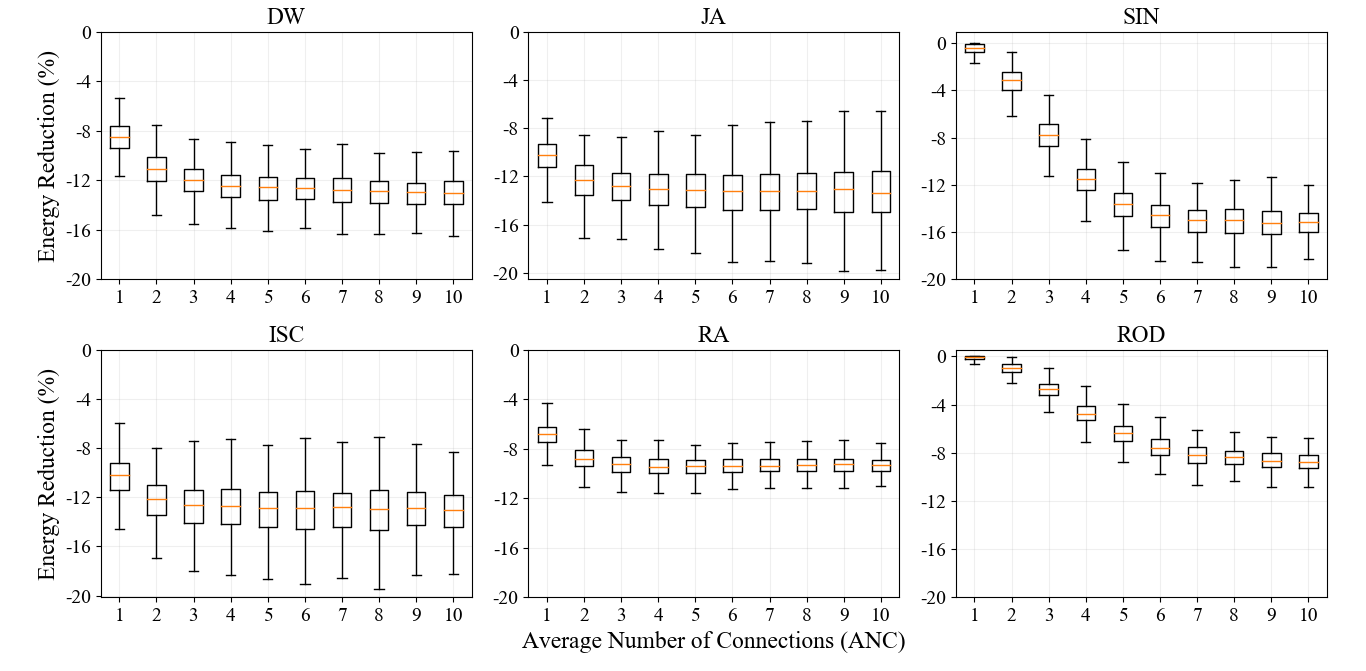}
    \caption{Sensitivity analysis of OD models to ANC}
    \label{fig:sens}
\end{figure}

In the previous section for each run, a random number from 1 to 10 has been selected for ANC.  In this section the sensitivity of simulation outputs to this factor have been investigated for each OD model. The results are surmised in Figure \ref{fig:sens}.

It is obvious that increasing the number of relationships in the community can immediately increase the likelihood of being connected to low energy consumers, which can lead to higher energy savings. However, the amount of savings with increasing ANC remains limited. In this example, the amount of energy savings has increased approximately as long as ANC reaches to the number of five.
As expected, not all models are sensitive to ANC in the same way. SIN and ROD have shown the greatest sensitivity, whilst other OD models are less affected. This observation could be explained by the weight factors in SIN and ROD's dependence on ANC.

As the final comment, a summary of the simulation study's findings and their implications is provided:
\begin{itemize}
    \item The choice of OD model for eco-feedback simulation is critical, since different OD models can affect the result by up to 50\%.
    \item Among the five standard OD models used in this study, the Relative Agreement (RA) model predicted more accurately on the basis of field observations on energy savings derived from feedback interventions. Furthermore, SIN has demonstrated the greatest sensitivity to the average number of connections (ANC).
    \item The Revised OD model (ROD), a hybrid OD model, produced the most accurate findings in comparison. This means that considering the closeness of energy behaviour as well as the closeness of social ties while modelling energy behaviour modification through eco-feedback interventions could improve the model's prediction.
\end{itemize}

\section{Conclusion}
Occupancy behaviour has a significant impact on building energy usage and emissions. As a result, occupancy energy interventions such as feedback systems are provided as a suitable method of improving occupant behaviours and controlling energy demand. Because empirical studies involve a significant amount of time and effort, researchers frequently employ simulation models for their analyses. However, developing an appropriate model capable of coherently presenting real-world problems is difficult.

This paper examined five standard Opinion Dynamic (OD) models (DW, JA, SIN, RA, and ISC) for modelling eco-feedback programs, using feedback experiment reports as the benchmark, and discovered that the RA model produces substantially superior outcomes. A hard influence condition has been considered for interactions in this model, which restricts the number of behaviour changes and energy savings. Following a thorough examination of the aforementioned OD models, ROD model is proposed, showing remarkable accuracy. The weight-factor and influence condition of interactions have been changed in ROD, and both closeness of energy behaviour and social network have been taken into account.

Finally, there are several limitations to this article that could be examined further in future research. A model comparison was carried out in a residential setting for the short-term outcomes of eco-feedback interventions. However, the model's behaviour may differ in different contexts (commercial, for example) or in a long-term impact. Furthermore, the more detailed the model's inputs, the more accurate the model. Thus, future research could give realistic techniques for obtaining and quantifying occupancy characteristics (flexibility factors, social connections, and weight factors, for example) and their possible relationships.






\bibliographystyle{Other/model1-num-names.bst}
\bibliography{Main.bib}







\end{document}